\documentclass[aps,prd,showpacs,preprintnumbers,nofootinbib,floatfix,floats,groupedaddress,twocolumn]{revtex4}

\usepackage{bm}
\usepackage{latexsym}
\usepackage{dcolumn}
\usepackage{amsmath,amsfonts,amssymb}
\usepackage{graphicx,epsfig}
\usepackage{color}
\usepackage[active]{srcltx}
\usepackage{subfig}
\usepackage{slashed}
\usepackage{tikz}
\usetikzlibrary{shapes,snakes}
\def\nn{\nonumber}

\def\l{\left}
\def\r{\right}
\def\DM{\mathrm{d}}

\reversemarginpar

\def \lp {\ell_0}

\def \Rsq {\widetilde {\rm \bf Ric}(p,p_0)}
\def \Rsqn {\widetilde {\rm \bf Ric}}
\def \Rsg {{\rm \bf Ric}(p_0)}
\def \sigG {\Sigma_{G, p_0}}

\newcommand{\boxedeqn}[1]
{
\begin{equation}
  \addtolength{\fboxsep}{5pt}
   \boxed{
   \begin{aligned}
      #1
   \end{aligned}
   }
\end{equation}
}

\begin{document}

\title{Small scale structure of spacetime: van Vleck determinant and {\it equi-geodesic} surfaces}

 \author{D. Jaffino Stargen}
 \email{jaffino@physics.iitm.ac.in}
 \affiliation{Department of Physics, Indian Institute of Technology Madras, Chennai 600 036}

 \author{Dawood Kothawala}
 \email{dawood@physics.iitm.ac.in}
 \affiliation{Department of Physics, Indian Institute of Technology Madras, Chennai 600 036}

\date{\today}
\begin{abstract}
\noindent It has recently been argued that if spacetime $\mathcal M$ possesses non-trivial structure at small scales, an appropriate semi-classical description of it should be based on non-local bi-tensors instead of local tensors such as the metric $g_{ab}(p)$. Two most relevant bi-tensors in this context are Synge's World function $\Omega(p,p_0)$ and the van Vleck determinant (VVD) $\Delta(p,p_0)$, as they encode the metric properties of spacetime and (de)focussing behaviour of geodesics. They also characterize the leading short distance behavior of two point functions of the d'Alembartian $_{p_0} \square_p$. 
\\
\\
We begin by discussing the intrinsic and extrinsic geometry of {\it equi-geodesic} surfaces $\sigG \equiv \{ p \in \mathcal M| \Omega(p,p_0)= \rm constant \}$ in a geodesically convex neighbourhood of an event $p_0$, and highlight some elementary identities relating the VVD with geometry of $\sigG$. As an {\it aside}, we also comment on the contribution of $\sigG$ to the surface term in the Einstein-Hilbert (EH) action and show that it can be written as a {\it volume} integral of $\square \ln \Delta$.
\\
\\
We then proceed to study the small scale structure of spacetime in presence of a Lorentz invariant short distance cut-off $\lp$ using $\Omega(p,p_0)$ and $\Delta(p,p_0)$, based on some recently developed ideas. We derive a $2$nd rank bi-tensor $q_{ab}(p,p_0;\lp)=q_{ab}\l[ g_{ab}, \Omega, \Delta \r]$ which naturally yields geodesic intervals bounded from below and reduces to $g_{ab}$ for $\Omega \gg \lp^2/2$. We present a general and mathematically rigorous analysis of short distance structure of spacetime based on (a) geometry of {equi-geodesic surfaces} $\sigG$ of $g_{ab}$, (b) structure of the non local d'Alembartian $\widetilde{ _{p_0} \square_p}$ associated with $q_{ab}$, and (c) properties of VVD. In particular, we prove the following: (i) The Ricci {\it bi-scalar} $\Rsq$ of $q_{ab}$ is completely determined by $\sigG$, the tidal tensor and first two derivatives of $\Delta(p,p_0)$, and has a non-trivial {\it classical} limit (see text for details): 
$$\lim \limits_{\lp \rightarrow 0} \lim \limits_{\Omega \rightarrow 0^{\pm}} \Rsq = \pm D R_{ab} q^a q^b$$ 
(ii) The GHY term in EH action evaluated on equi-geodesic surfaces straddling the causal boundaries of an event $p_0$ acquires a non-trivial structure.
\\
\\
These results strongly suggest that the mere existence of a Lorentz invariant minimal length $\lp$ can leave {\it unsuppressed residues} independent of $\lp$ and (surprisingly) independent of many precise details of quantum gravity. For e.g., the coincidence limit of $\Rsq$ is finite as long as the modification of distances $\mathcal S_{\lp}: 2\Omega \mapsto 2 \widetilde \Omega$ satisfies (i) $\mathcal S_{\lp}(0)=\lp^2$ (the condition of minimal length), (ii) $\mathcal S_{0}(x)=x$, and (iii) $\l[ |\mathcal S_{\lp}| / \mathcal S_{\lp}'^2 \r](0) < \infty$. In particular, the function $\mathcal S_{\lp}(x)$, which should eventually come from a complete framework of quantum gravity, need not admit a perturbative expansion in $\lp$.
\\
\\
Finally, we elaborate on certain technical and conceptual aspects of our results in the context of entropy of spacetime and classical description of gravitational dynamics based on Noether charge of Diff invariance instead of the EH lagrangian. 
\end{abstract}

\maketitle
\vskip 0.5 in
\noindent
\maketitle
\section{Introduction} \label{sec:intro} 
Quantum effects are expected to drastically affect the structure of space and time at the smallest of scales. However, our current theories of gravity and quantum mechanics are (fortunately) very stingy with the options they leave us as far as the small scale of structure of spacetime is concerned. For example, attempts to model such a structure by violating or deforming Lorentz invariance (LI) are either very strongly constrained by experiments, or run into deeper conceptual issues when one goes beyond the simple one-particle models. It is much more plausible that instead of LI, it is the assumption of locality that might have to be given up at small scales \cite{locality}. However, abandoning locality then also necessitates that we give up the classical description of spacetime in terms of local tensorial objects, in particular the metric tensor $g_{ab}(p)$. Finding the right geometric variables that can describe spacetime geometry down to smallest scales is of utmost significant not only for quantum 
gravity, but also for 
the proper physical interpretation of the field equations of gravity at the classical level. In particular, the deep connection between Einstein equations, thermodynamics, and information theory that has been studied in depth for over a decade very strongly suggests that we question the conventional description of gravitational dynamics based on Einstein-Hilbert (EH) action. In any case, to properly understand the implications of results that have been accumulated from the study of quantum fields in curved spacetime, it is extremely important that one must first identify the correct geometric variables to describe spacetime geometry at the classical level itself. 

Fortunately, the hint for doing so also comes from these very same results. In particular, one of the most significant results to have come out of semiclassical studies is the existence of a minimal spacetime length, say $\lp$, below which spacetime intervals loose any operational significance \cite{ml-reviews, paddy-zero-point-length}. Such a {\it zero-point length} appears in various forms in several candidate models of quantum gravity, and is often considered as the universal regulator for divergences in quantum field theory (QFT) and general relativity (GR). In a recent work, one of us \cite{dk-ml} proposed that a more appropriate description of spacetime geometry in presence of a minimal length scale must be based on non-local bi-tensors instead of the metric tensor. The geodesic distance between two spacetime events, in particular, was proposed as a more fundamental object than the metric tensor. The relevant bi-tensor in this context is the so called Synge World function $\Omega(p,p_0)$ defined by \cite{synge-book}
\begin{eqnarray}
\Omega(p,p_0) &=& \frac{1}{2} \l( \lambda(p) - \lambda(p_0) \r) \int \limits_{\lambda(p_0)}^{\lambda(p)} \l[g_{ab} q^a q^b\r](x(\lambda)) \DM \lambda
\nn \\
&=& \frac{1}{2} \sigma(p,p_0)^2
\end{eqnarray}
where $\sigma(p,p_0)^2$ is the square of geodesic interval, with the corresponding geodesic distance given by $$d(p,p_0) = \sqrt{\epsilon \sigma(p,p_0)^2}$$ Here, $q^a$ is tangent to the geodesics, and $\epsilon=q^a q_a=\pm 1$. (In this paper, we shall use $\Omega(p,p_0), \sigma(p,p_0)^2$ and $d(p,p_0)$ interchangeably to keep the expressions and notation convenient. We will also often use $\lambda=d(p,p_0)$ in covariant Taylor series to keep track of terms of various orders without messing up the notation.)

Assuming that the small scale structure of spacetime is characterised (at least at a semiclassical level) by the existence of a minimal length, it was shown that one can construct a second rank bi-tensor, $q_{ab}(p,p_0;\lp)$, such that it yields geodesic distances with a lower bound $\lp$. 

The construction of $q_{ab}$ in \cite{dk-ml} was based on two inputs: 

{\bf P1:} The requirement that geodesic distances have a Lorentz invariant lower bound {\it and} this arises from modification of geodesic distances as $\sigma^2 \rightarrow \sigma^2 + \lp^2$. 

{\bf P2:} The requirement that the modified d'Alembartian $\widetilde{ _{p_0} \square_p}$ yields the following modification for the two point functions $G(p,p_0)$ of fields in {\it flat spacetime}: $G\l[\sigma^2\r] \rightarrow \widetilde G\l[\sigma^2\r] = G\l[ \sigma^2 + \lp^2 \r]$. (This essentially regulates the UV divergences in QFT, and is based on several earlier works on the subject.) 

These two requirements then completely fix the form of $q_{ab}$. Being manifestly covariant, the extension to arbitrary curved spacetimes suggests itself naturally.  

In subsequent work \cite{dk-tp}, it was shown that the Ricci bi-scalar $\Rsq$ corresponding to this so called {\it qmetric} $q_{ab}$ has a very specific non-analytic structure which results in a non-trivial result for the coincidence limit $\l[\Rsqn\r]$ when $\lp \rightarrow 0$. The specific result proved there was 
\begin{eqnarray}
\lim \limits_{\lp \rightarrow 0} \lim \limits_{\sigma^2 \rightarrow 0^{\pm}} \Rsq \propto R_{ab} q^a q^b
\end{eqnarray}
with $q^a$ being arbitrary normalised vectors at each spacetime event. A similar analysis for the surface term $K \sqrt{|h|}$ of the EH action was also presented subsequently, and the very same term as above was shown to appear there as well (see second reference in \cite{dk-tp}).

The above results have many deep implications, in particular for understanding better the notion of entropy associated with each spacetime event $p_0$ and it's causal boundaries, and for the {\it emergent gravity paradigm}; these were discussed in detail in \cite{dk-tp}. They not only provide a very strong hint towards the importance of the quantity $R_{ab} q^a q^b$ in description of gravitational dynamics, but also give a precise quantitative manner in which the transmutation of gravitational lagrangian $ R \rightarrow R_{ab} q^a q^b $ can arise as a relic of a minimal length. 

On the other hand, the analysis presented in \cite{dk-tp} does not give much insight on the robustness of the conclusions drawn from the final result, and much less insight on some of the miraculous cancellations responsible for it. In particular, the following issues concerning the main inputs {\bf P1} and {\bf P2} were left unclear:

1. The analysis {\it assumed} (see {\bf P1}) that a lower bound on geodesic distances is realized via the modification $\sigma^2 \rightarrow \sigma^2 + \lp^2$. While the motivation for such a modification comes from several older results, it remained unclear as to how much the final result depends on it. This question is of fundamental significance, since the precise manner in which a minimal length is introduced in spacetime can come only from a complete framework of quantum gravity. In absence of such a framework, it is important not to make any assumptions on how distances can get modified. In particular, the modifications introduced by quantum gravity can be non-perturbative, and hence need not possess a series expansion in $\lp$ near $\sigma^2=0$.

In this paper, we shall establish our result without making any such assumption. Technically, we shall keep the function $\mathcal S_{\lp}: 2\Omega \mapsto 2 \widetilde \Omega$, which represents modification of distances, completely arbitrary and satisfying only $\l[ |\mathcal S_{\lp}| / \mathcal S_{\lp}'^2 \r](0) < \infty$ in addition to it's defining properties (which will be given below). In particular, the function $\mathcal S_{\lp}$ need not admit a perturbative expansion in $\lp$, unlike the form $\mathcal S_{\lp}(x) = x + \lp^2$ which was used in \cite{dk-ml}. The construction of $q_{ab}$ for arbitrary $\mathcal S_{\lp}(x)$ was already sketched in Appendix A of \cite{dk-tp}, except for a crucial difference, which brings us to our second point concerning the input {\bf P2}.

2. The requirement {\bf P2} that two point functions get modified as $G\l[\sigma^2\r] \rightarrow \widetilde G\l[\sigma^2\r] = G\l[ \sigma^2 + \lp^2 \r]$ makes sense only when $G(p,p_0)$ depends on $p$ and $p_0$ only 
through $\sigma^2$ {$\forall (p,p_0)$}. This can not happen in arbitrary curved spacetimes, which very much reduces the possibilities available to fix the qmetric. This is, of course, good, since it reduces the room available for {\it adhoc} choices. The most general space(time)s in which $G(p,p_0)$ is only a function of $\sigma^2$ {$\forall (p,p_0)$} are the maximally symmetric spaces, of which flat space(time) is but the simplest possibility with zero curvature. Fixing the qmetric based on flat spacetime, although it captures the correct leading singularity of the two point functions in the coincidence limit, wipes away all information about curvature. More precisely, the leading singular structure of two point functions associated with the d'Alembartian ${}_{p_0} \square_p$ in arbitrary spacetime is given by the Hadamard form 
\begin{eqnarray}
G(p,p_0) := \frac{\sqrt{\Delta}}{\l(\sigma^2\r)^{\frac{D-2}{2}}} \times \l( 1 + \mathrm{smooth~terms} \r)
\end{eqnarray}
As is evident, the information about curvature, at the leading order, therefore appears in the two point functions through the so called van Vleck determinant (VVD) $\Delta(p,p_0)$. In flat spacetime, $\Delta(p,p_0)=1$ exactly, whereas in arbitrary curved spacetimes, 
\begin{eqnarray}
\lim \limits_{p \rightarrow p_0} \Delta(p,p_0) = 1
\end{eqnarray}
One might therefore think that the dependence of qmetric on $\Delta(p,p_0)$ can not possibly affect the coincidence limit of the Ricci biscalar $\Rsq$ (or any other curvature invariant) associated with $q_{ab}$. This expectation is, however, wrong. Curvature involves second derivatives of the metric, and the coincidence limit of the second (or higher) derivatives of $\Delta(p,p_0)$ is not zero in general. (The exact form of the coefficients in covariant Taylor expansion of VVD are well known, and are quoted later in this paper.) This makes it crucial to identify the dependence of qmetric on VVD. As we shall show, doing so leads to some remarkable results, all following from certain identities satisfied by the VVD.

3. The modified $\Rsq$ derived in \cite{dk-tp} is in fact singular in the coincidence limit $\sigma^2 \rightarrow 0$. This divergence is cubic in $q^a$'s, and can be regularised using known methods in point splitting regularisation, as was suggested in \cite{dk-tp}. However, this still leaves a certain amount of discomfort at the mathematical level. It does not make much sense to appeal to point splitting regularisation since our starting point, based on existence of a minimal length, does not invoke point splitting at any level, but is instead based on use of a non local second rank bi-tensor. It is therefore important to have a deeper look at this divergence and it's origin, particularly so because it depends on $\nabla_i R_{ab}$ and hence vanishes for all maximally symmetric spaces! The argument given in 2 above advocating the use of maximally symmetric spaces therefore is not expected to help here, since the divergence is anyway zero for these spaces. One requires a much more mathematically rigorous 
analysis to probe the structure of this divergent term. As we shall show, identifying the correct dependence of $q_{ab}$ on the VVD in fact cancels this divergence in a rather surprising manner. In fact, the reason for this cancellation is buried deep within the expansion of extrinsic curvature of {\it equi-geodesic surfaces} $\sigG$ (see below) in an arbitrary spacetime to $4$th order in covariant Taylor series, and a close relationship between VVD and extrinsic curvature of $\sigG$.

We address all the above issues in this paper, and while doing so, reveal the mathematical robustness of the final result, and hence it's inevitability (given the two basic inputs) in any theory which admits a LI short distance cut-off. Our key inputs would be much less restrictive and/or specialized than the ones used in \cite{dk-ml, dk-tp}, which makes the results presented here significantly stronger. These can be stated as:

{\bf Q1:} The requirement that geodesic distances have a Lorentz invariant lower bound. 

{\bf Q2:} The requirement that the modified d'Alembartian $\widetilde{ _{p_0} \square_p}$ yields the following modification for the two point functions $G(p,p_0)$ of fields in {\it all maximally symmetric spacetimes}: $G\l[\sigma^2\r] \rightarrow \widetilde G\l[\sigma^2\r] = G\l[ {\mathcal S}_{\lp}\l[ \sigma^2 \r] \r]$.

As is evident, these inputs are much more minimalistic compared with {\bf P1, P2}, and hence can be expected to provide much more general insights into small scale structure of spacetime.

\begin{figure*}%
    \centering
    \subfloat[Equi-geodesic surfaces $\sigG$ attached to an event $p_0$ in an arbitrary curved spacetime.]{{\includegraphics[width=0.6\textwidth]{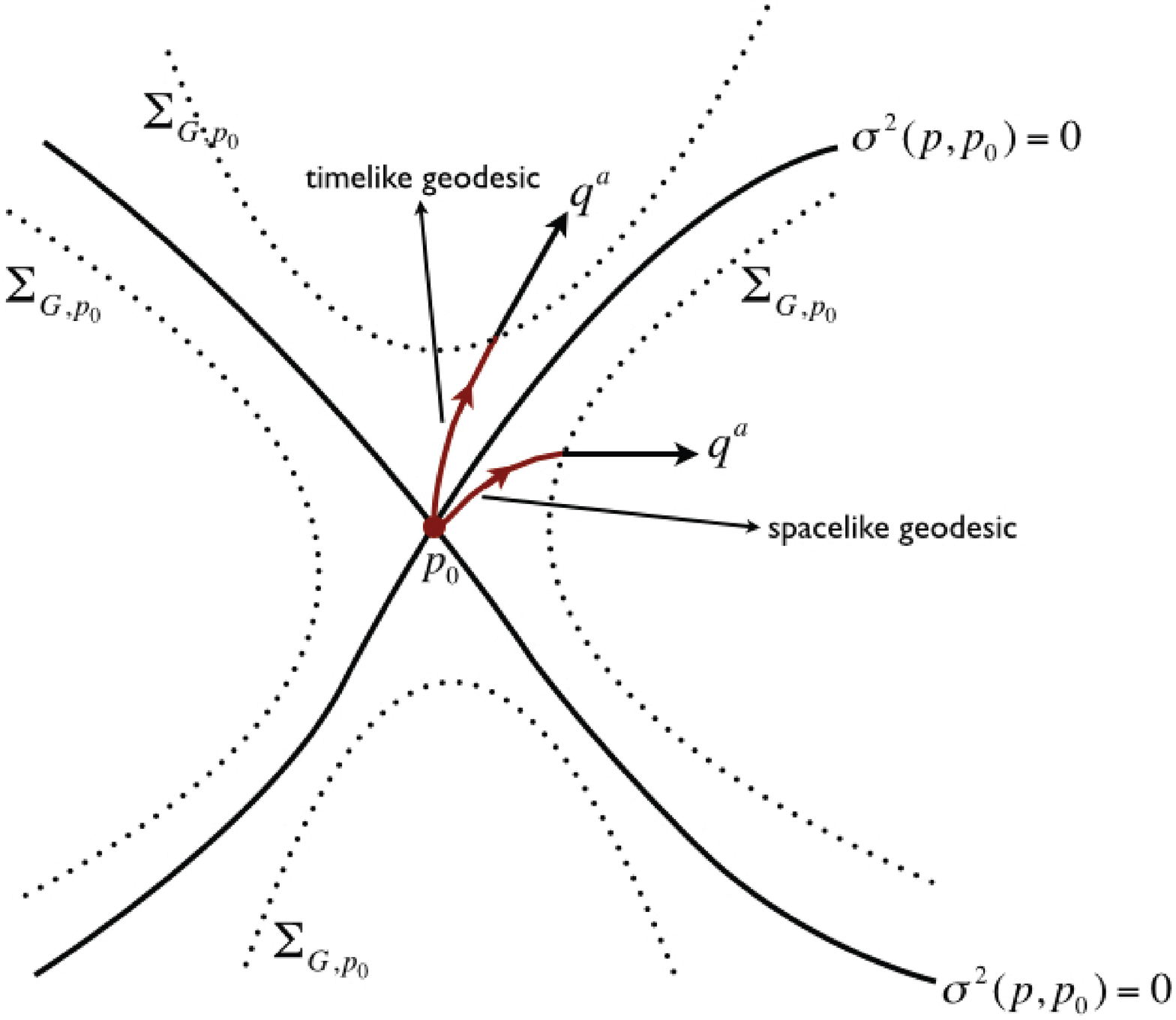} }}%
    \subfloat[$\sigG$ in Minkowski spacetime.]{{\includegraphics[width=0.325\textwidth]{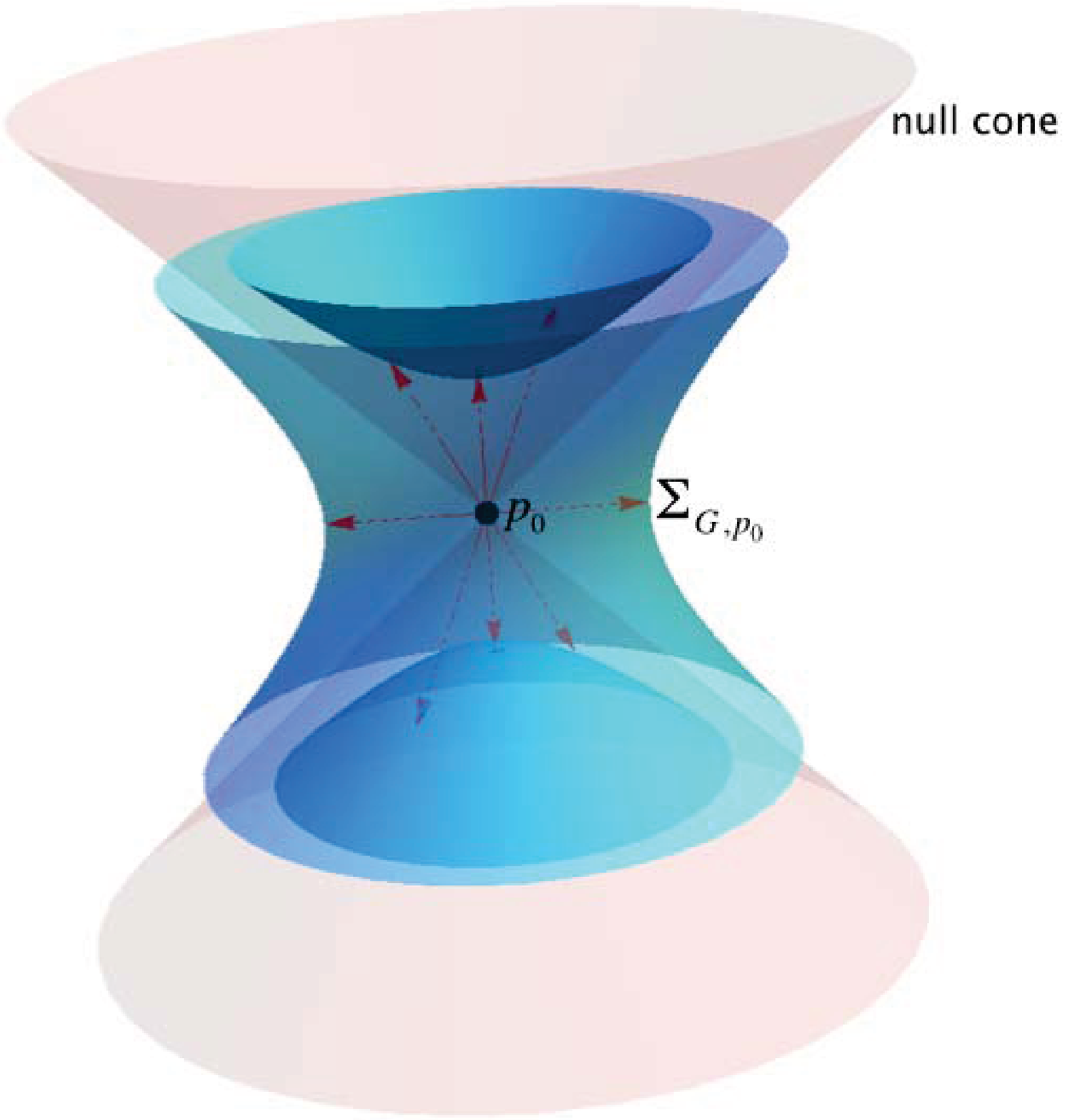} }}%
    \caption{The geodesic structure of spacetime.}%
    \label{fig:geodesic-congruence}%
\end{figure*}

The paper is structured as follows: 

Sec. \ref{sec:VVD}: In this section, we discuss the geodesic structure of arbitrary curved space(time)s, with focus on intrinsic and extrinsic geometry of equi-geodesic surfaces $\sigG$ \cite{dk-grg}, which comprise of points $p$ which are at some constant geodesic distance $\sigma^2$ from $p_0$, and connected to $p_0$ by non-null geodesics. We also discuss the geometric significance of VVD in studying the small scale structure of spacetime, and highlight some elementary identities relating derivatives of VVD to the extrinsic curvature of $\sigG$, which are used later in Sec. \ref{sec:rs}. 

Sec. \ref{sec:qmetric}: We present the derivation of the 2nd rank bi-tensor, the qmetric $q_{ab}(p,p_0;\lp)$, based on the two inputs {\bf Q1, Q2} stated above. In particular, we identify the dependence of the qmetric on the VVD using our condition {\bf Q2}.

Sec. \ref{sec:rs}: The Ricci bi-scalar $\Rsq$ for the qmetric is obtained in a closed form based on certain tools developed in \cite{dk-grg}, and it's coincidence limit $\sigma^2 \rightarrow 0$ is evaluated to obtain a local scalar $\l[\Rsqn\r](p_0)$ at $p_0$. It is then shown $\l[\Rsqn\r](p_0) \neq {\rm \bf Ric}(p_0)$, which is one of the key results of this paper. 

Sec. \ref{sec:ksqrth}: In this section, we complete our analysis of the EH action by evaluating the Gibbons-Hawking-York (GHY) surface term in the action on equi-geodesic surfaces, for the qmetric. 

In Sec. \ref{sec:discussion}, we finally conclude with a general discussion and implications of the results obtained in this work.

\underline{\textit{Key equations}:} 

The key results of this paper are contained in the boxed Eqs. (\ref{eq:qmfinal}), (\ref{eq:finalRsq}) and (\ref{eq:grin}).

\underline{\textit{Notation}:} 

We work in $D$ dimensions, and use the sign convention $(-, +, +, \ldots)$ for Lorentzian spaces. Latin alphabets denote spacetime indices. Also, for notational convenience, we use $\lp^2$ throughout to denote short distance cutoff on geodesic distances; for timelike/spacelike cases, the replacement $\lp^2 \rightarrow \epsilon \lp^2$ must be made in the final results after which $\lp^2>0$. For convenience, we give below a quick list of some of the most recurring symbols/notation used in the text:
\\
\\
$\bullet$ $D_k \overset{\rm def}{: \Longrightarrow} D-k$
\\
$\bullet$ $\mathcal E_{ab} = R_{ambn} q^m q^n$ 
\\
$\bullet$ $\mathcal E = g^{ab} \mathcal E_{ab}$
\\
$\phantom{\bullet}$ $\phantom{\mathcal E} = R_{ab} q^a q^b$
\\
$\bullet$ $\l[\Rsqn\r](p_0)$ is the coincidence limit of $\Rsq$

\section{The geodesic structure of spacetime} \label{sec:VVD} 
%
\subsection{Equi-geodesic surfaces}
%
Mathematically, a key role in our analysis would be played by the congruence of geodesics emanating from a fixed spacetime event $p_0$ and the surface comprised of events $p$ lying at constant geodesic interval from $p_0$, which we call as the {\it equi-geodesic} surface of event $p_0$ and denote it by $\sigG$. The relevant geometrical properties of such surfaces in arbitrary curved spacetimes were discussed in \cite{dk-grg}, and we simply quote the results which we will need here.


We start with the affinely parametrized tangent vector $q^a$ to the geodesic connecting $p_0$ to $p$
\begin{eqnarray}
q_a = \frac{\nabla_a \sigma^2}{2 \sqrt{\epsilon \sigma^2}}
\end{eqnarray}
and note that it is also the normal to $\sigG$. The extrinsic curvature tensor of $\sigG$, is therefore given by
\begin{eqnarray}
K_{ab} &=& \nabla_a q_b
= \frac{\nabla_a \nabla_b \l( \sigma^2/2 \r) - \epsilon q_a q_b}{\sqrt{\epsilon \sigma^2}}
\end{eqnarray}
This particular foliation, which characterizes the local geodesic structure of any spacetime, has many interesting properties, and all of these derive from the well known covariant Taylor series expansion of the bi-tensor 
$\nabla_a \nabla_b \l( \sigma^2/2 \r)$ at $p$ near $p_0$ \cite{christenson-worldf}:
\begin{eqnarray}
\nabla_a \nabla_b \l( \frac{1}{2} \sigma^2 \r) &=& g_{ab} - \frac{\lambda^2 }{3} \mathcal{E}_{ab} + \frac{\lambda^3}{12} \nabla_{\bm q} \mathcal{E}_{ab} 
\nn \\
&-& \frac{\lambda^4}{60} \l(\nabla^2_{\bm q} \mathcal{E}_{ab} + \frac{4}{3} \mathcal{E}_{ia} \mathcal{E}^i_{\phantom{i}b} \r) + O(\lambda^5)
\nn \\
\end{eqnarray}
where $\nabla_{\bm q} \equiv q^i \nabla_i$.

Therefore, we see that the extrinsic geometry of such a equi-geodesic ``foliation" is very special, and completely characterized by the {\it tidal tensor} $\mathcal E_{ab} = R_{a m b n} q^m q^n$. In fact, the intrinsic and extrinsic curvatures can be characterized by systematic Taylor expansions around $p_0$, given by \cite{dk-grg}
\begin{eqnarray}
K_{ab} &=& \frac{1}{\lambda}h_{ab} - \frac {1} {3} \lambda \mathcal{E}_{ab} + \frac {1} {12} \lambda^2 \nabla_{\bm q} \mathcal{E}_{ab} - \frac {1} {60}\lambda^3 F_{ab} + O(\lambda^4)
\nn \\
\nn \\
K &=& \frac{D_1}{\lambda} - \frac {1} {3} \lambda \mathcal{E} + \frac {1} {12} \lambda^2 \nabla_{\bm q} \mathcal{E} - \frac {1} {60}\lambda^3 F + O(\lambda^4)
\nn \\
\nn \\
\mathcal R_{\sigG} 
&=& \frac {\epsilon D_1 D_2} {\lambda^2} + R - \frac{2\epsilon(D+1)}{3} \mathcal{E} + O(\lambda)
\label{eq:taylor-exp}
\end{eqnarray}
where $F_{ab} = \nabla_{\bm q}^2 \mathcal{E}_{ab} + (4/3) \mathcal{E}_{ak} \mathcal{E}^k_{\phantom{k}b}$, and $F = F_{ab} g^{ab}$. For later use, we also quote here the combination (easily derived from above): 
\begin{widetext}
\begin{eqnarray}
K_{ab}^2 - \eta K^2 = \l(1 - \eta D_1\r) \Biggl\{ \frac{D_1}{\lambda^2} - \frac{2}{3} \mathcal{E} + \frac{1}{6} \lambda \nabla_{\bm q} \mathcal{E} - \frac{1}{30} \lambda^2 \l( \nabla_{\bm q}^2 \mathcal{E} - \frac{4}{3} \mathcal{E}_{ab}^2 \r) \Biggl\} \; + \; \frac{1}{9} \lambda^2 \l( \mathcal{E}_{ab}^2 - \eta \mathcal{E}^2 \r) + O(\lambda^3)
\label{eq:kabcomb}
\end{eqnarray}
\end{widetext}
for any arbitrary $\eta$. As we shall see, the structure of the above expression, which requires keeping upto $4$th order terms in $K_{ab}$ and $K$ (that is, terms of $O(\lambda^3)$), hold the key to elimination of coincidence limit divergences in $\Rsq$, in conjunction with a couple of differential identities (involving $K_{ab}$ and $K$) satisfied by the VVD, which we discuss next.

\subsection{van Vleck determinant}
%
The van Vleck determinant, $\Delta(p,p_0)$, is an extremely important object in semi-classical physics. Geometrically, this bi-scalar governs the properties of geodesic congruences emanating from a point, say $p_0$, as a function of an arbitrary point $p$. The immense physical importance of this object certainly warrants a longer discussion than presented here, and we refer the reader to \cite{christenson-worldf, vvd-refs, avramidi} for the same. In fact, as we shall see, the geometrical significance of VVD holds the key to its relevance for the small scale structure of spacetime, a theme that will resonate constantly throughout this paper.

The VVD is defined as follows:
\begin{eqnarray}
\Delta(p,p_0) = \frac{1}{\sqrt{g(p) g(p_0)}} {\rm det} \l\{ \nabla^{(p)}_a \nabla^{(p_0)}_b \frac{1}{2} \sigma(p,p_0)^2 \r\}
\end{eqnarray}
Two of the most important differential identities that we shall use, connecting the VVD with the extrinsic curvature of $\sigG$, are the following: 
\begin{eqnarray}
I1:&& \hspace{0.1cm} \nabla_{\bm q} \ln \Delta = \frac{D_1}{\sqrt{\epsilon \sigma^2}} - K
\\
I2:&& \hspace{0.1cm} \nabla_{\bm q} \nabla_{\bm q} \ln \Delta = - \frac{D_1}{\epsilon \sigma^2} + K_{ab}^2 + R_{ab} q^a q^b
\label{eq:VVD-ids}
\end{eqnarray}
where $\nabla_{\bm q} \equiv q^i \nabla_i$ and $K_{ab}^2 \equiv K_{ab} K^{ab}$.

{\it Proofs of $I1$ and $I2$}: The above elementary identities follow trivially from the expression \cite{vvd-refs, avramidi, poisson-lrr}
\begin{equation}
 \nabla_i \l[\Delta \nabla^i \sigma^2 \r] = 2 D \Delta
\end{equation}
Noting that the acceleration $a^i$ of $q^i$ is zero since $q^i$ represents tangents to geodesics, we can write the above identity as
\begin{eqnarray}
\hspace{0.5cm} \nabla_{\bm q} \ln \Delta = \frac{D_1}{\sqrt{\epsilon \sigma^2}} - K
\end{eqnarray} 
which is $I1$. Operating once more with $\nabla_{\bm q}$, and using the (easily proved) differential geometric identity 
\begin{eqnarray}
\nabla_{\bm q} K &=& q^i \nabla_i \nabla_j q^j
\nn \\
&=& - K_{ab}^2 - R_{ab} q^a q^b + \nabla_i a^i
\end{eqnarray}
with $a^i=0$, we get $I2$.

\subsection{ Aside: VVD and the surface term in Action}
\label{sec:vvd-surfaceterm}
%
As an aside, let us point out the possible relevance of the VVD in the gravitational action when one focusses on an observer dependent description of gravitational dynamics based on causal structure associated with an arbitrary 
event (``observer") $p_0$. The relevance of such a description has gained increased attention since the proposal by Jacobson \cite{jacobson-eos} of using local Rindler frames as probes of gravitational dynamics. We shall focus on the equi-geodesic surfaces $\sigG$ straddling the causal boundaries of an arbitrary event $p_0$, and briefly comment on the null limit in the end. 

The complete EH action is given by \cite{toolkit}
\begin{eqnarray}
16 \pi \mathcal A_{EH} = \int \limits_{\mathcal V} R \, \DM V_D + 2 \epsilon \int \limits_{\partial \mathcal V} \l(K - K_0 \r) \DM \Sigma_{D-1}
\end{eqnarray}
where $\DM V_D, \DM \Sigma_{D-1}$ are covariant volume elements in bulk and boundary respectively. The subtraction term, $K_0$, is usually taken to be the trace of extrinsic curvature of the boundary surface embedded in flat spacetime. Usually, the boundary $\partial V$ is taken at infinity. However, given the fact that the causal structure of spacetime limits the amount of information accessible at an event $p_0$, it is interesting to ask for the contribution of the boundaries $\sigG$ (which, in the null limit, would make the null cone of $p_0$). We therefore write
\begin{eqnarray}
16 \pi \mathcal A_{EH} = \int \limits_{\mathcal V_{p_0}} R \, \DM V_D + 2 \epsilon \l( \int \limits_{\partial \mathcal V_{\infty}} + \int \limits_{\sigG} \r) \l(K - K_0 \r) \DM \Sigma_{D-1} \nn
\end{eqnarray}
(Note the subscript $p_0$ on $\mathcal V$; we put it as a reminder that we are now focussing on quantities from the point of view of a specific event (observer) $p_0$.)

The trace of extrinsic curvature of $\sigG$, as embedded in {\it flat} spacetime, is $K_0 = D_1/\sqrt{\epsilon \sigma^2}$. Recalling $I1$, this immediately implies 
$$
\l( K - K_0 \r)_{\sigG} = - q^i \nabla_i \ln \Delta
$$
Using divergence theorem, $\int_{\sigG} q^i \nabla_i \ln \Delta = \int_V \square \ln \Delta$. (We must include a similar contribution from $\partial \mathcal V_{\infty}$; we do not write this explicitly since we are only interested in the contribution from $\sigG$). Putting all this together, we get
\begin{eqnarray}
16 \pi \mathcal A_{EH} = \int \limits_{\mathcal V_{p_0}} \l( R - 2 \epsilon\, \square \ln \Delta \r) \DM V_D + \mathcal A_{\partial \mathcal V_{\infty}}
\end{eqnarray}
where we have dumped all contributions from $\partial \mathcal V_{\infty}$ in $\mathcal A_{\partial \mathcal V_{\infty}}$. Let us now comment briefly on the null limit. Since the term $\square \ln \Delta$ is purely geometrical, we expect the null limit of the bulk term above to be straightforward. However, we must point out that the issue of boundary term for null boundaries is not completely unambiguous \cite{tp-null-counterterm}, and it might therefore require more care to repeat the above steps for a strictly null surface. 

The above analysis strongly suggests that an observer dependent study of gravitational dynamics might require us to change the conventional description based on EH lagrangian. In the rest of this paper, we will actually present a much stronger result suggesting a very natural transmutation of gravitational lagrangian from $R$ to $R_{ab} q^a q^b$ in presence of a Lorentz invariant short distance cut-off.
\section{The qmetric} \label{sec:qmetric} 
We now have the basic geometric tools using which we can implement {\bf Q1} and {\bf Q2} to arrive at a geometrical description of spacetime at small scales. Our aim in this section would be to construct
the so called {\it qmetric} $q_{ab}(p,p_0;\lp)$, as described in \cite{dk-ml}, which would reduce to the background spacetime metric $g_{ab}(p)$ for $d(p,p_0) \gg \lp$, but which yields geodesic distances bounded from below 
by $\lp$, while maintaining Lorentz invariance. As was shown in \cite{dk-ml}, the general form of $q_{ab}$ turns out to be (throughout this paper, $q^a=g^{ab} q_b$)
\begin{eqnarray}
\label{qu}
q^{ab} &=& A^{-1} g^{ab} + \epsilon Q q^a q^b
\nn \\
&=& A^{-1}h^{ab} + \epsilon (A^{-1}+Q) q^a q^b
\end{eqnarray}
with corresponding covariant components $q_{ab}$ 
\begin{eqnarray}
\label{ql}
q_{ab} &=& A g_{ab} - \epsilon B q_a q_b
\end{eqnarray}
where $B \equiv {QA}/\l({A^{-1}+Q}\r)$, where $h^{ab} = g^{ab} - \epsilon q^a q^b$ is the induced metric on $\sigG$, and $A, Q$ are functions of events $p, p_0$ to be fixed by {\bf Q1} and {\bf Q2}. 

The requirement of minimal length, {\bf Q1}, can be imposed \cite{dk-ml, dk-tp} using the Hamilton-Jacobi equation satisfied by $\sigma^2=2\Omega$ \cite{poisson-lrr}
\begin{eqnarray}
g^{ab} \partial_a \sigma^2 \partial_b \sigma^2 = 4 \sigma^2
\end{eqnarray}
and requiring
\begin{eqnarray}
q^{ab} \partial_a \mathcal S_{\lp} \partial_b \mathcal S_{\lp} = 4 \mathcal S_{\lp}
\label{eq:hjq}
\end{eqnarray}

We make no assumptions about precisely how quantum gravity would actually affect geodesic intervals, that is, we construct $q_{ab}$ for {\it arbitrary} modification of distances 
$\mathcal S_{\lp}: 2 \Omega \rightarrow 2 \widetilde \Omega$. We will only require:

(i) $\mathcal S_{\lp}(0)=\lp^2$ (the condition of minimal length).
\\
(ii) $\mathcal S_{0}$ is identity: $\mathcal S_0(2\Omega)=2\Omega$. 
\\
(iii) $\l[ |\mathcal S_{\lp}| / \mathcal S_{\lp}'^2 \r](0) < \infty$.

The Hamilton-Jacobi equation, Eq.~(\ref{eq:hjq}), then partially fixes the following combination in the qmetric, as was sketched in appendix A of \cite{dk-tp}:
\begin{equation}
\alpha \equiv A^{-1}+Q = \frac{1}{\sigma^2} \frac{\mathcal S_{\lp} (\sigma^2)}{\mathcal S_{\lp}'^2(\sigma^2)}
\label{eq:fix1}
\end{equation}
We now use {\bf Q2} to fix the qmetric completely (which is where we differ significantly with the presentation in \cite{dk-ml, dk-tp}). Recalling the condition {\bf Q2} explained in detail in the Introduction, we will require that the two point functions of the modified d'Alembartian $\widetilde{ _{p_0} \square_p}$ satisfy $\widetilde G\l[\sigma^2\r] = G\l[\mathcal S_{\lp}(\sigma^2)\r]$ in all maximally symmetric spacetimes (rather than just flat spacetime).

We start with the d'Alembartian operator corresponding to $q_{ab}$ for arbitrary backgrounds $g_{ab}$ (not necessarily maximally symmetric). After some algebra, we obtain:
\begin{widetext}
\begin{eqnarray}
\label{boxq}
\widetilde \square = A^{-1} \Biggl\{ \square_g+\frac{1}{2}D_3~g^{ij}\partial_i\ln A~\partial_j+\epsilon\slashed{\partial}\ln A~\slashed{\partial} \Biggl\}
+ \epsilon Q \Biggl\{ \l[ \nabla_i q^i + \frac{1}{2}D_1\slashed{\partial}\ln A\r]\slashed{\partial}+\slashed{\partial}^2 \Biggr\}
+ \sqrt{\epsilon \sigma^2}\alpha'\slashed{\partial}
\end{eqnarray}
\end{widetext}
where $D_k\equiv D-k$, $\slashed{\partial}\equiv q^i\partial_i$. 

To impose {\bf Q2}, we will analyse this operator for maximally symmetric spacetimes, in which $A$ and $Q$ are functions of only $\sigma^2$, and Eq.~(\ref{boxq}) becomes
\begin{equation}
\label{boq}
\widetilde \square = \alpha \; \square + 2\alpha\sigma^2\l[\ln \l(\alpha A^{D_1}\r)\r]'\frac{\partial}{\partial\sigma^2}
\end{equation}
On the other hand, the d'Alembertian $\square$ for maximally-symmetric spacetimes is given by
\begin{equation}
\label{bosi}
\square = \frac{\partial^2}{\partial \sigma^2}+\l(\frac{\partial}{\partial \sigma}\ln \Delta^{-1}+\frac{D_1}{\sigma}\r)\frac{\partial}{\partial \sigma}
\end{equation}
where 
$$\Delta^{-1/(D-1)} = \Biggl\{ \frac{\sin(|\sigma|/a)}{|\sigma|/a}, 1, \frac{\sinh(|\sigma|/a)}{|\sigma|/a} \Biggl\}$$ 
is the exact expression of the van Vleck determinant in maximally symmetric spacetimes of positive, zero, and negative curvature respectively (with radius of curvature $a$). The quantity $\Delta_{\mathcal S}$ below is defined as above with 
$\sigma^2 \rightarrow \mathcal S_{\lp}$.

We are now ready to impose {\bf Q2}. As shown in the Appendix, the condition that $\widetilde{G}\l[ \sigma^2 \r] = G\l[\mathcal S_{\lp}(\sigma^2)\r]$ is the two point function corresponding to $\widetilde \square$, which translates into 
$\widetilde \square \widetilde{G}\l[ \sigma^2 \r]=0$ when $\square G\l[ \sigma^2 \r] = 0$ (for $p\neq p_0$), gives a differential equation
\begin{equation}
 \frac{\DM}{\DM\sigma^2}\ln\l(\frac{A}{\mathcal S_{\lp}/\sigma^2}\l(\frac{\Delta_{\mathcal S}}{\Delta_{\phantom{\mathcal S}}}\r)^{2/D_1}\r)=0
\end{equation}
whose solution is 
\begin{eqnarray}
A = \frac{\mathcal S_{\lp}}{\sigma^2} \l( \frac{\Delta_{\phantom{\mathcal S}}}{\Delta_{\mathcal S}} \r)^{2/D_1}
\label{eq:fix2}
\end{eqnarray}
where the constant of integration is fixed by the condition $A=1$ when $\mathcal S_{\lp}=\sigma^2$. 

We have therefore accomplished our aim of identifying the dependence of $A$, and hence the qmetric, on the VVD by appealing to maximally symmetric spaces and {\bf Q2}. 
Equations (\ref{eq:fix1}) and (\ref{eq:fix2}) fix the final form of the qmetric as
\begin{widetext}
\boxedeqn{
 \bm q &= \frac{\mathcal S_{\lp}}{\sigma^2} \l(\frac{\Delta_{\phantom{\mathcal S}}}{\Delta_{\mathcal S}}\r)^{+\frac{2}{D_1}} \bm g \; + \; 
 \epsilon \Biggl\{ \frac{\sigma^2 \mathcal S_{\lp}'^2}{S_{\lp}} - \frac{\mathcal S_{\lp}}{\sigma^2} \l(\frac{\Delta_{\phantom{\mathcal S}}}{\Delta_{\mathcal S}}\r)^{+\frac{2}{D_1}} \Biggl\} \; \bm {q \otimes q}
 \label{eq:qmfinal}
}
\end{widetext}
with the inverse metric given by
\begin{eqnarray}
q^{ab} &=& \frac{\sigma^2}{\mathcal S_{\lp}}\l(\frac{\Delta_{\phantom{\mathcal S}}}{\Delta_{\mathcal S}}\r)^{-\frac{2}{D_1}} g^{ab} + 
\nn \\
&& \hspace{1.5cm} \epsilon \Biggl\{ \frac{S_{\lp}}{\sigma^2 \mathcal S_{\lp}'^2} - \frac{\sigma^2}{\mathcal S_{\lp}}\l(\frac{\Delta_{\phantom{\mathcal S}}}{\Delta_{\mathcal S}}\r)^{-\frac{2}{D_1}} \Biggl\} q^a q^b
 \nn
\end{eqnarray}

For maximally-symmetric spacetimes, it can be shown that the metrics $g_{ab}$ and $q_{ab}$ are related by a non-local, singular diffeomorphism. This is most easily seen from the line element corresponding to $q_{ab}$ when $g_{ab}$ is maximally symmetric (assuming $\sigma^2>0$ and constant positive curvature below for purpose of demonstration):
		\begin{eqnarray}
		 \DM s^2 &=& \DM\sigma^2+\sigma^2 \Delta^{-2/D_1}\DM\Omega^2_{D-1}
		 \nn \\
		 \widetilde{\DM s^2} &=&q_{ab}\DM x^a\DM x^b
		 \nn \\
		 &=& \l( \DM \sqrt{\mathcal S_{\lp}} \r)^2 + \mathcal S_{\lp} \Delta_{\mathcal S}^{-2/D_1}\DM\Omega^2_{D-1}
		\end{eqnarray}
The above relationship between $g_{ab}$ and $q_{ab}$ will, of course, not hold in arbitrary curved spacetimes, and therefore the two metrics would have different curvatures.

The form derived in \cite{dk-ml} and \cite{dk-tp} turn out to be special cases of the one derived above if one chooses $\mathcal S_{\lp}(x)=x+\lp^2$ and $\Delta=1$. As we will see, while the choice 
of $\mathcal S_{\lp}$ is just that, a choice, putting $\Delta=1$ can be potentially dangerous, since one then risks missing important contributions to $\Rsq$ arising from derivatives of $\Delta$. 

In fact, this is just what happens. 

%
\section{Ricci scalar for the qmetric} \label{sec:rs} 
%
Having found the qmetric, Eq.~(\ref{eq:qmfinal}), in terms of modification of geodesic distances $\mathcal S_{\lp}$ and the VVD, we can now proceed to evaluate the Ricci {\it bi}-scalar $\Rsq$ corresponding to it. This is the simplest curvature invariant associated with any spacetime, and more importantly for us, the Ricci scalar is the simplest lagrangian describing gravitational dynamics in general relativity. We can then construct a scalar from $\Rsq$ by taking the coincidence limit $p \rightarrow p_0$, and compare it with $\Rsg$, the Ricci scalar of the background spacetime $g_{ab}$. Naively, one might expect that 
$$ \l[ \Rsqn \r](p_0) \overset{?}{=} \Rsg + {\rm terms~of~order~} \lp$$
We will explicitly calculate $\lim \limits_{\lp \rightarrow 0} \l[ \Rsqn \r](p_0)$ to verify this, and show that the leading term is $\neq \Rsg$. We will find an exact expression for the leading term as well as sub-leading 
terms in terms of the geometry of $\sigG$ and the first two derivatives of the VVD.

To proceed with the calculation, we use the following expression, derived in \cite{dk-grg}, relating Ricci scalars of metrics related in a manner similar to $q_{ab}$ and $g_{ab}$.
\begin{align}
\Rsq &= \Omega^{-2} \, \Rsg
\;+\; \epsilon \l( \alpha - \Omega^{-2} \r)  \mathcal{J}_d 
\;-\; \epsilon \alpha \, \mathcal{J}_c %
\nn \\
\label{eq:mRs-final}
\end{align}  
(where, borrowing notation of \cite{dk-grg}, $\Omega^2 = A$), and
\begin{eqnarray}
\mathcal{J}_c &=&
\epsilon \l[ 2 D_1 \Omega^{-1} \square \Omega + D_1 D_4 \Omega^{-2} (\nabla \Omega)^2 \r]
\nn 
\\
\vspace{.2cm}
&& \hspace{.55cm} \; + \l( K + D_1 \nabla_{\bm q} \ln \Omega \r) \times \nabla_{\bm q} \ln \alpha \Omega^2 
\nn \\
\nn \\
\mathcal{J}_d &=&
2 R_{ab} q^a q^b + K_{ab}^2 -K^2 
\nn \\
&=& \epsilon \l( R - \mathcal R_{\sigG} \r)
\label{eq:JcJd}
\end{eqnarray}
%
One can now plug in the form of $A$ and $\alpha$ from Eqs.~(\ref{eq:fix1}) and (\ref{eq:fix2}) to find the form of RHS. This is the most important, and also the most lengthy, part of the calculation. The computation is largely aided by identities $I1$ and $I2$ (Eqs.~(\ref{eq:VVD-ids})) satisfied by the VVD. Some of the key steps are sketched in the Appendix \ref{app:Rsdetail}.

The final result turns out to be
\begin{widetext}
\boxedeqn{
\label{eq:finalRsq}
      \Rsq &= \underbrace{
      \Biggl[ \frac{\sigma^2}{\mathcal S_{\lp}} \zeta^{-{2}/{D_1}} \; \mathcal R_{\sigG} - \frac{D_1 D_2}{\mathcal S_{\lp}} + 4 (D+1) (\ln \Delta_{\mathcal S})^{\bullet} \Biggl]
      }_{\bm{Q_0}}
			\;-\; 
			\underbrace{\frac{\mathcal S_{\lp}}{\lambda^2 \mathcal S'^2_{\lp}}  \Biggl\{ K_{ab} K^{ab} - \frac{1}{D_1} K^2 \Biggl\}}_{\bm{Q_{\rm K}}}
			\\
			&\phantom{\Biggl[ \frac{\sigma^2}{\mathcal S_{\lp}} \zeta^{-{2}/{D_1}} \; \mathcal R_{\sigG} \;-\; \frac{D_1 D_2}{\mathcal S_{\lp}} + 4 (D+1) \overset{\bullet}{\Delta}_{\mathcal S} \Biggl] \;-\; \;\;\;\;\;}
			+ \underbrace{4 \mathcal S_{\lp}  \Biggl\{ - \frac{D}{D_1} \l[ (\ln \Delta_{\mathcal S})^{\bullet} \r]^2 + 2 (\ln \Delta_{\mathcal S})^{\bullet\bullet} \Biggl\}}_{\bm{Q_\Delta}}
}
\end{widetext}
where we have defined $\zeta={\Delta}/{\Delta_{\mathcal S}}$, $(\ln \Delta_{\mathcal S})^{\bullet} = \DM \ln \Delta_{\mathcal S}/\DM \mathcal S_{\lp}$, and 
$(\ln \Delta_{\mathcal S})^{\bullet\bullet} = \DM(\ln \Delta_{\mathcal S})^{\bullet}/\DM \mathcal S_{\lp}$. It is not too difficult to see that for $\lp=0, \mathcal S_0(x)=x$, the RHS above reduces to $\Rsg$. (To verify this, 
one has to use $I1 ,I2$ from Eqs.~(\ref{eq:VVD-ids}) above along with $\nabla_{\bm q} \equiv 2 \epsilon \lambda \DM/\DM \sigma^2$.) 

It is crucial to note here that we have not used any of the covariant Taylor expansions yet; the above expression, therefore, does not assume the region of spacetime under consideration to be smooth (i.e., having finite curvature). This will be important for discussing the implications of our framework for cosmological and black hole singularities, which we wish to address in future work. 

For the purpose of this paper, however, we shall assume that we are working in smooth regions of spacetime, so that the various Taylor expansions given in Sec.~\ref{sec:VVD} can be used. The significance of separating out the RHS into $\bm{Q_0, Q_{\rm K}}$ and $\bm{Q_\Delta}$ will become evident shortly.

The above expression holds the key to understand non-perturbative effects of a covariant short distance cut-off on spacetime curvature. Let us therefore first highlight some of it's most important mathematical aspects, before taking it's $\lp \rightarrow 0$ limit.

1. The expression contains no derivatives of the function $\mathcal S_{\lp}(x)$ higher than one. This is an extremely delicate mathematical point; as can be seen from the details provided in the Appendix, terms of the form $\mathcal S''_{\lp}$ do in fact appear in the intermediate steps, but they cancel out in the final expression. Since $\mathcal S_{\lp}(x)$ represents (in general non-perturbative) effects of quantum gravity on invariant distance between spacetime events, the non-existence of higher derivatives of $\mathcal S_{\lp}$ in $\Rsq$ is of deep conceptual importance -- it tells us that semi-classical effects of quantum gravity can be captured only via limited information about the precise details of quantum gravity.

2. The Ricci bi-scalar $\Rsq$ is completely described by {\it geodesic structure of spacetime}, characterized by: 
\\
(a) $\mathcal R_{\sigG}$ \hspace{.25cm} (intrinsic curvature)
\\
(b) $K_{ab}$ \hspace{.7cm} (extrinsic curvature), and
\\
(c) the van Vleck determinant $\Delta(p,p_0)$.

3. The extrinsic curvature of $\sigG$ appears only in a very special combination, which (as we will see in a moment), is responsible for {\it no coincidence limit divergences}! This is essentially a consequence of 
Eq.~(\ref{eq:kabcomb}) for $\eta=1/D_1$.

\subsection{$\lim \limits_{\lp \rightarrow 0} \l[\Rsqn\r](p_0)$}

The coincidence limit of $\Rsq$ gives us a local scalar $\l[\Rsqn\r](p_0)$ at each spacetime event $p_0$ which will depend on $\lp$. We wish to ask whether this scalar gives back $\Rsg$, 
the Ricci scalar of the background spacetime, when $\lp$ is set to zero.

The limit $\lp \rightarrow 0$ must be taken with care. First of all, note that any $\lp$ independent contribution must come from $\bm{Q_0}$. The contribution from $\bm{Q_{\rm K}}$ and $\bm{Q_\Delta}$ can only be $O(\lp^2)$, since $\mathcal S_{\lp}(0)=\lp^2$. 

Let us therefore first focus on $\bm{Q_0}$.

To do this, we invoke the coincidence limit expansions of various quantities given in Eqs.~(\ref{eq:taylor-exp}), in addition to the following well known
covariant Taylor expansion of the VVD
\begin{eqnarray}
\Delta^{1/2}(p,p_0) = 1 + \frac{1}{12} \lambda^2 R_{ab} q^a q^b + O(\lambda^3)
\end{eqnarray} 
from which it is easy to see that 
\begin{eqnarray}
\lim \limits_{\lp \rightarrow 0} \lim \limits_{\sigma^2 \rightarrow 0} (\ln \Delta_{\mathcal S})^{\bullet} = \frac{1}{6} \epsilon \l[ R_{ab} q^a q^b \r](p_0)
\label{eq:Q0step1}
\end{eqnarray} 
Also, using the last of Eqs.~(\ref{eq:taylor-exp}) for $\mathcal R_{\sigG}$, and the fact that $\Delta(0)=1$, we get
\begin{eqnarray}
\lim \limits_{\sigma^2 \rightarrow 0}  \Biggl\{ \frac{\sigma^2}{\mathcal S_{\lp}} \zeta^{-\frac{2}{D_1}} \mathcal R_{\sigG} - \frac{D_1 D_2}{\mathcal S_{\lp}} \Biggl\} = \frac{D_1 D_2}{\mathcal S_{\lp}(0)} \l( \Delta^{\frac{2}{D_1}}_{\lp} - 1 \r)
\nn 
\end{eqnarray} 
where $\Delta^{1/2}_{\lp} = 1 + \frac{1}{12} \epsilon \lp^2 \l[ R_{ab} q^a q^b \r](p_0) + \ldots$ 

The limit $\lp \rightarrow 0$ limit of the RHS above is most easily evaluated using the {\it l'Hospital}'s rule (note that both the numerator and denominator vanish in this limit):
\begin{eqnarray}
\lim \limits_{\lp \rightarrow 0} \frac{D_1 D_2}{\mathcal S_{\lp}(0)} \l( \Delta^{2/D_1}_{\lp} - 1 \r) &=& \lim \limits_{\lp \rightarrow 0} \frac{D_1 D_2}{\partial_{\lp^2} \mathcal S_{\lp}(0)} \partial_{\lp^2} \Delta^{2/D_1}_{\lp}
\nn \\
&=& \frac{1}{3} D_2 \epsilon \l[ R_{ab} q^a q^b \r](p_0)
\label{eq:Q0step2}
\end{eqnarray} 

From Eqs.~(\ref{eq:Q0step1}, \ref{eq:Q0step2}), we immediately get
\begin{eqnarray}
\lim \limits_{\lp \rightarrow 0} \lim \limits_{\sigma^2 \rightarrow 0} {\bm Q_0} &=& \epsilon \l[ \frac{D-2}{3} + \frac{4(D+1)}{6} \r] \l[ R_{ab} q^a q^b \r](p_0)
\nn \\
&=& \epsilon D \l[ R_{ab} q^a q^b \r](p_0)
\nn \\
&=& \epsilon D \mathcal E(p_0)
\end{eqnarray}
\\
which is one of the most important results in this paper. The above limit, being independent of $\mathcal S_{\lp}(x)$, is precisely the relic left by the presence of a zero point length. (Qualitatively, this is similar to the various quantum anomalies one encounters in QFT in curved spacetimes \cite{birrel}; see \cite{dk-tp} (first reference) for a much detailed conceptual discussion of this and related points.)

Now consider $\bm{Q_{\rm K}}$, with the condition 
$\l[ |\mathcal S_{\lp}| / \mathcal S_{\lp}'^2 \r](0) < \infty$. This term would be divergent in the coincidence limit were it not for the fact that the combination of extrinsic curvature appearing here is $O(\lambda^2)$, as is readily seen from Eq.~(\ref{eq:kabcomb}) with $\eta=1/D_1$. This is a very strong result. Any other combination of extrinsic curvature tensor would lead to coincidence limit divergences in $\l[\Rsqn\r](p_0)$, and the reason the right combination appears here is completely due to the presence of VVD. In the absence of it, we would indeed get divergences, as can be seen from the result in \cite{dk-tp}. 

The final term, $\bm{Q_\Delta}$, is a smooth function which would yield further $O(\lp^2)$ dependent terms coupled to the background curvature. These terms can be read off from the known covariant Taylor expansions of the VVD.

To summarize, then, we have proved the following:
\boxedeqn{
\label{eq:grin}
\lim \limits_{\lp \rightarrow 0} \l[ \Rsqn \r](p_0) &=& \epsilon D \l[ R_{ab} q^a q^b \r](p_0)
}

$\bm \star$ \underline{\textit{Aside}:} {The expansion in powers of $\lp$}.

For the sake of completeness, we quote below the higher order terms in $\lp$ which can be obtained if one further assumes that all the quantities involved allow a legitimate series expansion in $\lp$ in the coincidence limit. We must, however, caution that such an assumption could be highly objectionable (even wrong) in full quantum gravitational context. Nevertheless, the expansion turns out to be:
\begin{eqnarray}
\l[ \Rsqn \r](p_0) &=& \epsilon D {\mathcal E}(p_0) + \frac{2 \epsilon (D+1)}{3}(\nabla_{\bm q} {\mathcal E}) \lp
 \nn \\
&+& \Biggl\{\frac{1}{18}\biggl(D+2-\frac{2}{\mathcal S'(0)^2}\biggr)\l({\mathcal E}_{ab}^2 - \frac{{\mathcal E}^2}{D_1}\r) 
\nn \\
&+& \frac{1}{4}(D+2)\nabla_{\bm q}^2 {\mathcal E} \Biggl \} \lp^2+O(\lp^3)
\end{eqnarray}
\section{The surface term on $\sigG$} \label{sec:ksqrth} 
%
In the previous section, we analyzed the Ricci bi-scalar $\Rsq$ for the qmetric and showed that it's coincidence limit becomes proportional to $R_{ab}q^a q^b$ in the limit ${\lp \rightarrow 0}$. We now calculate the contribution of the surface term $\widetilde{K \sqrt{|h|}}$ for the equi-geodesic surfaces using the qmetric. 
This is a much simpler calculation than that of $\Rsq$, and is along the same lines as the one given in \cite{dk-tp} (second reference), so we will be brief.

We start with the following relation between induced metrics and the extrinsic curvatures \cite{dk-grg}
\begin{eqnarray}
\sqrt{|\widetilde h|} &=& A^{D_1/2} \sqrt{|h|}
\nn \\
\widetilde K &=& \sqrt{\alpha} \l(K+\frac{D_1}{2}\nabla_q\ln A\r)
\end{eqnarray}
Since by definition $K=\partial \ln \sqrt{|h|}/\partial \lambda$, the series for $K$ in Eq.~(\ref{eq:taylor-exp}) readily yields 
\begin{equation}
 \sqrt{|h|}=\lambda^{D_1} \l(1-\frac{1}{6} \mathcal E(p_0) \lambda^2 + \frac{1}{36} \nabla_{\bm q} \mathcal E (p_0) \lambda^3 + O(\lambda^4) \r)
\end{equation}
Putting everything together, we get
\begin{eqnarray}
\label{ksqh}
\lim_{\sigma^2 \rightarrow 0} \widetilde{K \sqrt{|h|}} = \frac{\lp^D}{\Delta_{\lp}} \l(\frac{D_1}{\lp^2} - 2 \lim \limits_{\sigma^2 \rightarrow 0} (\ln \Delta_{\mathcal S})^{\bullet}  \r)
\end{eqnarray}
The above limit of the surface term corresponds to the surfaces $\sigG$ straddling very close to the causal horizon of $p_0$. If one takes the limit $\lp \rightarrow 0$ above, it gives zero, and hence there is no non-trivial effect on the surface 
term as was found in $\Rsq$. 

$\bm \star$ \underline{\textit{Aside}:} {The expansion in powers of $\lp$}.

Again, {\it if one can justify a series expansion in $\lp$}, one gets
\begin{eqnarray}
 \lim_{\sigma^2 \rightarrow 0} \widetilde{K \sqrt{|h|}} &=& D_1 \lp^{D-2} \l[1 - \frac{D+1}{6 (D-1)}{\mathcal E}(p_0) \lp^2 + O(\lp^3) \r]
 \nn \\
 &\overset{D=4}{=}& 3 \lp^2 - \frac{5}{6}{\mathcal E}(p_0) \lp^4 + O(\lp^5)
\end{eqnarray}
in which the first two terms have the same form as in \cite{dk-tp} (second reference), where no expansion in $\lp$ was needed.

%

\section{Discussion} \label{sec:discussion} 
%
As mentioned in the Introduction, one needs only a very few conceptual inputs from semiclassical gravity to deduce some basic facts about spacetime at small scales. The existence of a lower bound on 
spacetime intervals is one such input, and if it turns out to be a fundamental feature of quantum gravity, it makes sense to look for a geometric description of spacetime in terms of objects which are likely to be more useful in incorporating such a lower bound. Bi-tensors, then, provide a natural choice, and indeed, if one looks deeper into our theories of classical GR and QFT, some of the most important features of these theories, such as geodesic deviation, focussing of geodesics, causal structure, singularity structure of two point functions etc, are characterized in terms of bi-tensors. Characterizing the small scale topology of spacetime is another aspect which would necessitate a description directly in terms of bi-tensors such as the distance function $d(p,p_0)$. This is the point of view which was stressed in \cite{dk-ml}, developed and described in much greater conceptual detail in \cite{dk-tp}, and has been put on a much more rigorous basis in the present paper. Our general expression (\ref{eq:finalRsq}) for the Ricci bi-scalar $\Rsq$ of the qmetric presents a natural basis for the description of gravitational dynamics by a non-local action. This can be particularly relevant for study of spacetime singularities, where one can not use covariant Taylor expansions. 
\footnote{The idea of gravity being described fundamentally by a non-local action, with geodesic distance playing the key role, seems to be {\it conceptually} in tune with certain ideas presented in DeWitt and Alvarez et. al. \cite{dewitt-alvarez-etal}.}

Although mathematical complexity forced us to look at only the Ricci scalar (instead of the full Riemann tensor) for geometries with a covariant short distance cut-off, the resulting expression very elegantly and concretely expresses the key idea: curvature of spacetime might be solely expressible in terms of behaviour of it's geodesics and related bi-tensorial quantites. In absence of a lower bound on geodesic distances, such a description would coincide with the standard one in terms of local tensors such as $g_{ab}(p), R_{abcd}(p)$ etc. However, if there exists a minimal length, then the non-local character of bi-tensors, combined with the non-analytic deformation of geometry necessitated by such a minimal length, might lead to a very different description of spacetime curvature at smallest of scales; in particular, it may leave a relic independent of the details or value of the short distance cut-off, thereby acting as a crucial guidepost towards our understanding of classical gravity itself. 

The mathematical results derived here, for example, seem to strongly support the so called emergent gravity paradigm, in which gravitational dynamics is described in terms of thermodynamics of future causal horizon of an event $p_0$. Two of the key ideas in this context -- the use of {\it local Rindler frames} as probes of spacetime curvature (due to Jacobson \cite{jacobson-eos}), and a variational principle based on {\it entropy functional} (due to Padmanabhan et. al. \cite{entropy-functional}) -- find a unified and purely geometric description in our framework in terms of equi-geodesic {\it surfaces} (which replace the Rindler {\it trajectories}) and the $\lp=0$ term of the coincidence limit of Ricci bi-scalar of the qmetric, which happens to have the same form as the entropy functional.

The possibility of description of geometry in terms of two point functions of quantum fields has also been emphasized, in a series of paper, by Kempf and collaborators \cite{kempf}. The connection with the work presented here is obvious: the UV behavior of two point functions is in one to one correspondence with vanishing of geodesic distances in the coincidence limit, which, of course, was the basis for our input {\bf Q2}. In fact, there could also be a more fundamental connection at the level of geometry itself. For example, it was pointed out in \cite{kempf-algebra} that the commutation relation between position and momenta, $\l[\hat{\bm x}^\mu, \hat{\bm p}_\nu\r]$, would in general acquire a correction on a curved manifold, thereby affecting the resultant uncertainties. We hope to present more details on this particular connection in a future work (which is in progress).

We hope to apply the results derived here to analyze implications of a Lorentz invariant minimal length for issues such as cosmological and black hole singularities, trans-planckian problem in black hole physics and cosmology, and possible relevance for the cosmological constant problem.
%

{\it Acknowledgements --} DK thanks T. Padmanabhan for comments and discussion on various aspects of the ideas presented here. DJS is supported by fellowship from the Ministry of Human Resource Development (MHRD), India.
\appendix 
\section{Derivation of Eq.~(\ref{eq:finalRsq})} \label{app:Rsdetail}

The complete derivation of Eq.~(\ref{eq:finalRsq}) is lengthy, but there are several key structural aspects of the result which are worth highlighting, since these are responsible not only for the final simple form of $\Rsq$, but also for 
\begin{itemize}
\item Finiteness of the coincidence limit, $\l[ \Rsqn \r](p_0)$ 
\item Cancellation of any derivatives of $\mathcal S_{\lp}(x)$ higher than the first.
\end{itemize}
The first of these is purely a consequence of the correct identification of the dependence of the qmetric on VVD, and repeatedly using the identities $I1, I2$ for the same. 

The second fact is more non-trivial, and there is no simple reason to have expected it! In fact, since $q_{ab}$ depends on $\alpha$, and $\alpha$ depends on $\mathcal S'_{\lp}$, one expects $\Rsq$ to involve terms such as $\mathcal S'''_{\lp}$, which it does not. Although we do not have a completely geometric understanding of why this {\it must} happen, we highlight below {\it how} the cancellations happen which point to the following two reasons for these subtle cancellations:
\begin{itemize}
\item The fact that the coupling between $q_{ab}$ and $g_{ab}$ is {\it disformal} rather than {\it conformal}.
\item The fact that the $\mathcal S''_{\lp}$ contribution of the conformal part is cancelled by a contribution coming from the disformal one.
\end{itemize}
To demonstrate these, we start from Eq.~(\ref{eq:mRs-final}), and note that the only possibility of occurrence of second derivatives of $\mathcal S_{\lp}$ is from $\mathcal J_c$, which we write as
\begin{eqnarray}
\mathcal{J}_c &=&
\underbrace{\epsilon \l[ 2 D_1 \Omega^{-1} \square \Omega + D_1 D_4 \Omega^{-2} (\nabla \Omega)^2 \r]}_{\mathcal{J}_{c1}}
\nn 
\\
\vspace{.2cm}
&& \hspace{.55cm} \; + \underbrace{\l( K + D_1 \nabla_{\bm q} \ln \Omega \r) \times \nabla_{\bm q} \ln \alpha \Omega^2 }_{\mathcal{J}_{c2}}
\end{eqnarray}
The cancellation of higher order derivatives happen between the terms $\mathcal{J}_{c1}$, which is the contribution of purely conformal transformation, and $\mathcal{J}_{c2}$, contributed by the disformal part. Simplifying the above expression, we get
\begin{eqnarray}
 \mathcal{J}_{c1} &=& -\frac{1}{\sqrt{\alpha}} \frac{D_1}{\sqrt{\epsilon S}} \nabla_q \ln \alpha+2 \epsilon \frac{\sqrt{\epsilon S}}{\sqrt{\alpha}} \nabla_q \ln \alpha~ (\ln \Delta_s)^{\bullet} \nonumber \\
 &+& \l( \mathrm{terms~that~do~not~depend~on~} \mathcal S''_{\lp} \r)
\end{eqnarray}
and
\begin{eqnarray}
 \mathcal{J}_{c2} &=& +\frac{1}{\sqrt{\alpha}} \frac{D_1}{\sqrt{\epsilon S}} \nabla_q \ln \alpha - 2 \epsilon \frac{\sqrt{\epsilon S}}{\sqrt{\alpha}} \nabla_q \ln \alpha~ (\ln \Delta_s)^{\bullet} \nonumber \\
 &+& \l( \mathrm{terms~that~do~not~depend~on~} \mathcal S''_{\lp} \r)
\end{eqnarray}
If the metrics $q_{ab}$ and $g_{ab}$ are conformally coupled, $\alpha=\Omega^{-2}$ and hence $\mathcal{J}_{c2}=0$. In such a case, the higher derivatives of $\mathcal S_{\lp}$ would not cancel; the cancellation is solely a consequence of the disformal coupling between the metrics.
\\
 


\end{document}